\documentclass[english,aps,preprint]{revtex4}
\usepackage[T1]{fontenc}
\usepackage[latin1]{inputenc}
\usepackage{amssymb}

\makeatletter

\newcommand{\noun}[1]{\textsc{#1}}

\usepackage{babel}
\makeatother
\begin{document}

\preprint{This line only printed with preprint option}

\title{Generalized Wick's theorem for multiquasiparticle overlaps as a limit
of Gaudin's theorem}

\author{Sara Perez-Martin}

\email{sara.perez@uam.es}

\affiliation{Departamento de Física Teórica C-XI, Universidad Autónoma de Madrid,
28049 Madrid, Spain}

\author{Luis M. Robledo}

\email{luis.robledo@uam.es}

\affiliation{Departamento de Física Teórica C-XI, Universidad Autónoma de Madrid,
28049 Madrid, Spain}

\begin{abstract}
By using the extension of the statistical Wick's theorem (Gaudin's
theorem) to deal with generalized statistical density operators (those
which can be expressed as the product of and operator carrying out
a canonical transformations times a density operator) and using the
appropriate limits we are able to rederive in a very simple way the
standard generalized Wick's theorem for overlaps of mean field wave
functions. Due to the simplicity of the derivation it is now straightforward
to consider more involved cases and some of them are discussed. The
present derivation also allows to obtain general and compact formulas
for other particular cases of the generalized Wick theorem involving
overlaps of multiquasiparticle excitations of product wave functions.
The new expressions allow to reduce the combinatorial complexity of
the standard calculation of the above overlaps.
\end{abstract}
\maketitle

\section{introduction}

Nowadays it is widely recognized the need for going beyond the mean
field approximation in order to better understand the low lying structure
of the atomic nucleus (see Ref \cite{Bender.03} for a thorough discussion
of the topic). The mesoscopic nature of the nucleus makes essential
the restoration of the symmetries broken at the mean field level in
order to incorporate into the wave function the expected behavior
under the symmetry's operators that are responsible for the selection
rules of many observable quantities. The correlation energy gained
by the restoration of symmetries and the consideration of quantum
fluctuations in relevant degrees of freedom plays a very relevant
role not only in the description of the nucleus' binding energy but
also in the distribution of states in the spectrum of the nucleus.
In the process of incorporating into the theoretical framework the
quantum fluctuations mentioned previously, tools like the generalized
Wick's theorem are widely used. The generalized Wick's theorem (GWT)
for fermions allows the evaluation of the overlap of a general product
of fermionic creation and annihilation operators between any two non
orthogonal product wave functions $|\Phi_{0}\rangle$ and $|\Phi_{1}\rangle$.
According to the GWT, the overlap $\langle\Phi_{1}|\hat{O}|\Phi_{0}\rangle/\langle\Phi_{1}|\Phi_{0}\rangle$,
where $\hat{O}$ is a general product of creation $\beta_{k}^{+}$
and annihilation $\beta_{k}$ quasiparticle operators, is given as
the sum of all possible contractions of all possible combinations
of two quasiparticle operators, namely \begin{eqnarray}
\overline{\mathbb{C}}_{11\quad mn} & = & \langle\Phi_{1}|\beta_{m}\beta_{n}|\Phi_{0}\rangle/\langle\Phi_{1}|\Phi_{0}\rangle\nonumber \\
\overline{\mathbb{C}}_{12\quad mn} & = & \langle\Phi_{1}|\beta_{m}\beta_{n}^{+}|\Phi_{0}\rangle/\langle\Phi_{1}|\Phi_{0}\rangle\label{eq:}\\
\overline{\mathbb{C}}_{21\quad mn} & = & \langle\Phi_{1}|\beta_{m}^{+}\beta_{n}|\Phi_{0}\rangle/\langle\Phi_{1}|\Phi_{0}\rangle\nonumber \\
\overline{\mathbb{C}}_{22\quad mn} & = & \langle\Phi_{1}|\beta_{m}^{+}\beta_{n}^{+}|\Phi_{0}\rangle/\langle\Phi_{1}|\Phi_{0}\rangle\nonumber \end{eqnarray}
These four contractions can be gathered together as members of the
bipartite matrix of contractions \begin{equation}
\overline{\mathbb{C}}=\left(\begin{array}{cc}
\overline{\mathbb{C}}_{11} & \overline{\mathbb{C}}_{12}\\
\overline{\mathbb{C}}_{21} & \overline{\mathbb{C}}_{22}\end{array}\right)\label{eq:}\end{equation}
that will prove to be a convenient quantity in future developments.
Both the proof of the generalized Wick's theorem as well as the explicit
expression of the contractions have been given by several authors
\cite{Balian.Brezin,Hara.79,Onishi.66} using different approaches.
In all the cases the proof of the theorem as well as the contractions
are obtained after quite lengthy considerations and in a manner that
is not prone to generalizations. On the other hand, the generalization
of the standard Wick's theorem for mean values to the framework of
an statistical admixture of quantum states (where mean values are
replaced by traces over the Fock space including a statistical density
operator) is rather straightforward and the proof of the theorem is
fairly simple \cite{Gaudin.60} owing to the simplifications introduced
by the cyclic property of the trace. The generalization of this theorem
(Gaudin's theorem thereafter) to the case where the statistical density
operator is replaced by a more general operator given by the product
of the density operator times an operator carrying out a canonical
transformation is also rather straightforward \cite{Rossignoli.Ring.94,Balian.Veneroni}.
The reason is that the spirit of the proof of Gaudin's theorem (the
cyclic invariance of the trace) is still valid and the proof of the
generalized theorem proceeds along the lines of the standard one.
The purpose of this paper is to show that the generalized Wick's theorem
can be deduced from the corresponding statistical version by an appropriate
limiting procedure of the {}``probabilities'' of the statistical
admixture. This limiting procedure together with the simplicity of
the statistical version of the theorem will allow us to derive formulas
for the most general situation in a fairly simple way. By generalizing
the previous limiting procedure we will be able to compute not only
the standard overlaps but also overlaps of multiquasiparticle excitations
that can be rather cumbersome to compute by using the standard techniques
\cite{Tanabe.99,Hara.95}. The novel expressions obtained in this
paper for the contractions are very compact and therefore their use
substantially simplifies the usually lengthy calculations (with a
complexity of combinatorial type) involved in considering those multiquasiparticle
overlaps.

\section{Definitions and proof of Gaudin's theorem}

\subsection{Basic definitions}

First of all let us introduce some basic notation that will be used
to simplify the expressions derived below (the notation closely follows
that of \cite{Balian.Brezin}). A given set of fermionic creation
and annihilation quasiparticle operators $\beta_{k}$ and $\overline{\beta}_{k}$
are defined in terms of a reference single particle creation and annihilation
set of operators $b_{l}$ and $\overline{b}_{l}$ by means of linear
combinations

\begin{equation}
\begin{array}{c}
\beta_{k}=\sum_{l}A_{lk}^{*}b_{l}+B_{lk}^{*}\overline{b}_{l}\\
\overline{\beta}_{k}=\sum_{l}C_{lk}b_{l}+D_{lk}\overline{b}_{l}\end{array}\label{eq:Bogoliubov}\end{equation}
The reference single particle creation $\overline{b}_{l}$ and annihilation
$b_{l}$ set of operators is assumed to satisfy canonical anticommutation
rules $\{ b_{l},b_{l'}\}=\{\overline{b}_{l},\overline{b}_{l'}\}=0$
and $\{\overline{b}_{l},b_{l'}\}=\delta_{ll'}$ but the operators
$b_{l}$ and $\overline{b}_{l}$ (as well as $\beta_{l}$ and $\overline{\beta}_{l}$)
are not necessarily related by hermitian conjugation. In the following,
creation and annihilation operators expressed with Latin letters will
correspond to single particle operators whereas the ones in Greek
letters will correspond to quasiparticles. For both kinds of operators
Latin subindices will be used. These and other relationships to be
considered below can be written in a more compact way by introducing
vectors of dimension $2N$ ($N$ is the total number of creation or
annihilation operators considered) incorporating both annihilation
and creation operators both for the quasiparticle\begin{equation}
\alpha_{\mu}=(\beta_{1},\beta_{2},\ldots,\beta_{N},\overline{\beta}_{1},\overline{\beta}_{2},\ldots,\overline{\beta}_{N})\label{eq:condensed}\end{equation}
and the single particle \begin{equation}
a_{\mu}=(b_{1},b_{2},\ldots,b_{N},\overline{b}_{1},\overline{b}_{2},\ldots,\overline{b}_{N})\label{eq:}\end{equation}
operators. For these vectors of operators Greek indices like the $\mu$
above are introduced. It is also convenient to introduce the notation\begin{equation}
\overline{\alpha}_{\mu}=(\overline{\beta}_{1},\overline{\beta}_{2},\ldots,\overline{\beta}_{N},\beta_{1},\beta_{2},\ldots,\beta_{N})\label{eq:}\end{equation}
as well as the $2N\times2N$ matrix\begin{equation}
\sigma=\left(\begin{array}{cc}
0 & 1\\
1 & 0\end{array}\right)\label{eq:sigma}\end{equation}
that allows us to express the $\overline{\alpha}_{\mu}$ set in terms
of $\alpha_{\mu}$ and viceversa as $\overline{\alpha}_{\mu}=\sum_{\nu}\sigma_{\mu\nu}\alpha_{\nu}$.
Now the linear combination of Eq. (\ref{eq:Bogoliubov}) can be written
in a more compact way as\begin{equation}
\alpha_{\mu}=\sum_{\nu}W_{\nu\mu}^{*}a_{\nu}\label{eq:BogoliubovCond}\end{equation}
where the $2N\times2N$ matrix $W$ has the following bipartite structure
\begin{equation}
W=\left(\begin{array}{cc}
A & C^{*}\\
B & D^{*}\end{array}\right)\label{eq:Wmatrix}\end{equation}
If it is required the new set of quasiparticle operators $\alpha_{\mu}$
to satisfy the canonical fermionic commutation relations \begin{equation}
\{\overline{\alpha}_{\mu},\alpha_{\nu}\}=\delta_{\mu\nu}\label{eq:}\end{equation}
 (and assuming that the single particle set $a_{\mu}$ does) then
the coefficients $W$ of the linear combination defining $\alpha_{\mu}$
in terms of $a_{\mu}$ must satisfy\begin{equation}
W\sigma W^{T}=\sigma\label{eq:CanonicalCond}\end{equation}
which is the condition for the transformation matrix $W$ to be a
canonical transformation. 

Nothing has been said about hermiticity since the operators $\overline{b}_{k}$
need not to be the hermitian conjugate of $b_{k}$ (the only relevant
property are the anticommutation relations). However, and in order
to simplify the following considerations we will assume in the following
that the single particle operators $b_{k}^{+}=\overline{b}_{k}$ are
the hermitian conjugate of the $b_{k}$. Concerning the quasiparticle
operators $\overline{\beta}_{l}$ we will assume that they are not
in general the hermitian conjugates of $\beta_{l}$. However, if we
require $\overline{\beta}_{l}=\beta_{l}^{+}$ to be the hermitian
conjugate of $\beta_{l}$ then the coefficients of the transformation
relating them to the set $b_{k}$ and $b_{k}^{+}$ must satisfy the
two requirements $A=D$ and $B=C$. In terms of the $W$ coefficients
this translates to $\sigma W\sigma=W^{*}$.

In order to prove the statistical Wick's theorem (or Gaudin's theorem)
mentioned in the introduction we only have to assume that there is
a statistical density operator $\hat{\rho}$, not necessarily hermitian,
and satisfying\begin{equation}
\alpha_{\mu}\hat{\rho}=\sum_{\nu}\mathbb{M}_{\mu\nu}\hat{\rho}\alpha_{\nu}\label{eq:aro}\end{equation}
where the matrix $\mathbb{M}$ is in principle arbitrary. However,
if this transformation is applied to the canonical commutation relation
$\hat{\rho}^{-1}\{\alpha_{\mu},\alpha_{\nu}\}\hat{\rho}=\sigma_{\mu\nu}$
we obtain $\{\hat{\rho}^{-1}\alpha_{\mu}\hat{\rho},\hat{\rho}^{-1}\alpha_{\nu}\hat{\rho}\}=\sigma_{\mu\nu}$
and using Eq. (\ref{eq:aro}) the condition $\mathbb{M}\sigma\mathbb{M}^{T}=\sigma$
is obtained for the matrix $\mathbb{M}$ implying that this matrix
indeed represents the one of canonical transformation (see Eq. (\ref{eq:CanonicalCond})).
Usually, the density operator as well as other operators carrying
out canonical transformations and to be used later are written in
terms of the exponential of one body operators (see Appendix A). In
this case the transformation matrix $\mathbb{M}$ is given by the
following exponential $\mathbb{M}=\exp(-\sigma\mathbb{K}_{A})$ where
$\mathbb{K}_{A}$ is a skew-symmetric matrix (see appendix A for details).
However, not all the matrices satisfying the canonical transformation
condition of Eq. (\ref{eq:CanonicalCond}) can be written as the above
exponential (the most general transformation in a space of dimension
2 is enough to find a counterexample, see appendix A) and in most
of the exposition we will consider the density operator $\hat{\rho}$
as a general operator not necessarily expressible as the exponential
of an one body operator. 

We also have to introduce the concept of the trace of an operator
$\hat{A}$ over the whole Fock space\begin{equation}
\mathrm{Tr}[\hat{A}]=\langle\phi|\hat{A}|\phi\rangle+\sum_{m}\langle\phi|\beta_{m}\hat{A}\beta_{m}^{+}|\phi\rangle+\frac{1}{2!}\sum_{mn}\langle\phi|\beta_{m}\beta_{n}\hat{A}\beta_{m}^{+}\beta_{n}^{+}|\phi\rangle+\cdots\label{eq:trace}\end{equation}
where $|\phi\rangle$ is a product wave function and $\beta_{m}$
and $\beta_{m}^{+}$ are the corresponding annihilation and creation
operators associated to $|\phi\rangle.$ The trace satisfies the cyclic
property $\mathrm{Tr}[\hat{A}_{1}\hat{A}_{2}\cdots\hat{A}_{K-1}\hat{A}_{K}]=\mathrm{Tr}[\hat{A}_{K}\hat{A}_{1}\hat{A}_{2}\cdots\hat{A}_{K-1}]$
that will turn out to be the fundamental property to prove Gaudin's
theorem.

\subsection{Proof of Gaudin's theorem}

Now let us consider the evaluation of the following trace\begin{equation}
\mathrm{Tr}[\hat{\rho}\alpha_{\mu_{1}}\alpha_{\mu_{2}}\cdots\alpha_{\mu_{K}}]\label{eq:}\end{equation}
with $K$ an even integer number. The result to be obtained is nothing
but Gaudin's theorem of Ref \cite{Gaudin.60} but it is repeated here
as it will be instructive for further generalizations to be considered
below. We start by moving $\alpha_{\mu_{1}}$ to the right by using
the anticommutation relations\begin{equation}
\mathrm{Tr}[\hat{\rho}\alpha_{\mu_{1}}\alpha_{\mu_{2}}\cdots\alpha_{\mu_{K}}]=\{\alpha_{\mu_{1}},\alpha_{\mu_{2}}\}\mathrm{Tr}[\hat{\rho}\alpha_{\mu_{3}}\alpha_{\mu_{4}}\cdots\alpha_{\mu_{K}}]-\mathrm{Tr}[\hat{\rho}\alpha_{\mu_{2}}\alpha_{\mu_{1}}\cdots\alpha_{\mu_{K}}]\label{eq:.}\end{equation}
and keep moving it to the right until we reach the last quasiparticle
operator in the argument of the trace\begin{eqnarray}
\mathrm{Tr}[\hat{\rho}\alpha_{\mu_{1}}\alpha_{\mu_{2}}\cdots\alpha_{\mu_{K}}] & = & \{\alpha_{\mu_{1}},\alpha_{\mu_{2}}\}\mathrm{Tr}[\hat{\rho}\alpha_{\mu_{3}}\alpha_{\mu_{4}}\cdots\alpha_{\mu_{K}}]\nonumber \\
 & - & \{\alpha_{\mu_{1}},\alpha_{\mu_{3}}\}\mathrm{Tr}[\hat{\rho}\alpha_{\mu_{2}}\alpha_{\mu_{4}}\cdots\alpha_{\mu_{K}}]\nonumber \\
 & + & \cdots+\nonumber \\
 & - & \mathrm{Tr}[\hat{\rho}\alpha_{\mu_{2}}\alpha_{\mu_{3}}\cdots\alpha_{\mu_{K}}\alpha_{\mu_{1}}]\label{eq:expr1}\end{eqnarray}
Using now the cyclic invariance of the trace the last term is written
as $\mathrm{Tr}[\alpha_{\mu_{1}}\hat{\rho}\alpha_{\mu_{2}}\alpha_{\mu_{3}}\cdots\alpha_{\mu_{K}}]$
and using the property of Eq. (\ref{eq:aro}) we arrive at\begin{equation}
\mathrm{Tr}[\hat{\rho}\alpha_{\mu_{2}}\alpha_{\mu_{3}}\cdots\alpha_{\mu_{K}}\alpha_{\mu_{1}}]=\sum_{\nu_{1}}\mathbb{M}_{\mu_{1}\nu_{1}}\mathrm{Tr}[\hat{\rho}\alpha_{\nu_{1}}\alpha_{\mu_{2}}\alpha_{\mu_{3}}\cdots\alpha_{\mu_{K}}]\label{eq:}\end{equation}
Moving now this term to the left hand side of Eq. (\ref{eq:expr1})
we obtain\begin{eqnarray}
\sum_{\mu_{1}}(\delta_{\mu_{1}\nu_{1}}+\mathbb{M}_{\nu_{1}\mu_{1}})\mathrm{Tr}[\hat{\rho}\alpha_{\mu_{1}}\alpha_{\mu_{2}}\cdots\alpha_{\mu_{K}}] & = & \{\alpha_{\nu_{1}},\alpha_{\mu_{2}}\}\mathrm{Tr}[\hat{\rho}\alpha_{\mu_{3}}\alpha_{\mu_{4}}\cdots\alpha_{\mu_{K}}]\label{eq:}\\
 & - & \{\alpha_{\nu_{1}},\alpha_{\mu_{3}}\}\mathrm{Tr}[\hat{\rho}\alpha_{\mu_{2}}\alpha_{\mu_{4}}\cdots\alpha_{\mu_{K}}]+\cdots\nonumber \end{eqnarray}
that can be written as\begin{equation}
\mathrm{Tr}[\hat{\rho}\alpha_{\mu_{1}}\alpha_{\mu_{2}}\cdots\alpha_{\mu_{K}}]=\mathbb{C}_{\mu_{1}\mu_{2}}\mathrm{Tr}[\hat{\rho}\alpha_{\mu_{3}}\alpha_{\mu_{4}}\cdots\alpha_{\mu_{K}}]-\mathbb{C}_{\mu_{1}\mu_{3}}\mathrm{Tr}[\hat{\rho}\alpha_{\mu_{2}}\alpha_{\mu_{4}}\cdots\alpha_{\mu_{K}}]+\cdots\label{eq:expr2}\end{equation}
where the matrix of contractions $\mathbb{C}_{\mu_{1}\mu_{2}}$ have
been introduced. They are given by\begin{equation}
\mathbb{C}_{\mu_{1}\mu_{2}}=\sum_{\nu_{1}}(1+\mathbb{M})_{\mu_{1}\nu_{1}}^{-1}\{\alpha_{\nu_{1}},\alpha_{\mu_{2}}\}=[(1+\mathbb{M})^{-1}\sigma]_{\mu_{1}\mu_{2}}\label{eq:contraction}\end{equation}
Now repeating the above procedure for the remaining traces of Eq.
(\ref{eq:expr2}) we arrive at the Gaudin's theorem which states that
the trace\begin{equation}
\frac{\mathrm{Tr}[\hat{\rho}\alpha_{\mu_{1}}\alpha_{\mu_{2}}\cdots\alpha_{\mu_{K}}]}{\mathrm{Tr}[\hat{\rho}]}\label{eq:}\end{equation}
equals the sum of the product of all possible contractions of Eq.
(\ref{eq:contraction}). It is important to point out here that the
above derivation is independent of the product wave function and quasiparticle
operators entering the definition of the trace in Eq. (\ref{eq:trace})
and the only relevant properties are the cyclic invariance of the
trace and the transformation law of Eq. (\ref{eq:aro}).

\subsection{Operator overlaps as limits of statistical traces}

Usually, the statistical density operator $\hat{\rho}$ is given as
the exponential $\hat{\rho}=\exp(-\hat{K})$ of an one-body hermitian
operator $\hat{K}=K^{0}+\frac{1}{2}\sum_{\mu\nu}\mathbb{K}_{\mu\nu}\alpha_{\mu}^{+}\alpha_{\nu}$
with \begin{equation}
\mathbb{K}=\left(\begin{array}{cc}
K^{11} & K^{20}\\
-K^{20\,*} & -K^{11\,*}\end{array}\right)\label{eq:}\end{equation}
The matrix $K^{11}$ is in this case hermitian whereas $K^{20}$ is
skew-symmetric. The matrix $\mathbb{M}$ corresponding to the representation
of the statistical density operator in the quasiparticle basis is
given by $\mathbb{M}=\exp(-\mathbb{K})$ and by working in the basis
where $\mathbb{K}$ is diagonal \begin{equation}
\mathbb{K}^{D}=\left(\begin{array}{cc}
k_{l} & 0\\
0 & -k_{l}\end{array}\right)\label{eq:}\end{equation}
the matrix $\mathbb{M}$ becomes diagonal\begin{equation}
\mathbb{M}=\left(\begin{array}{cc}
p_{l} & 0\\
0 & 1/p_{l}\end{array}\right)\label{eq:}\end{equation}
with $p_{l}=\exp(-k_{l})$. We will introduce the quantity $\rho_{\mu}=(p_{1},p_{2},\ldots,p_{N},1/p_{1},1/p_{2},\ldots,1/p_{N})$
such that $\mathbb{M}_{\mu\nu}=\rho_{\mu}\delta_{\mu\nu}$. The statistical
density operator can be now written as \begin{equation}
\hat{\rho}=|\phi\rangle\langle\phi|+\sum_{m}p_{m}\beta_{m}^{+}|\phi\rangle\langle\phi|\beta_{m}+\frac{1}{2!}\sum_{mn}p_{m}p_{n}\beta_{m}^{+}\beta_{n}^{+}|\phi\rangle\langle\phi|\beta_{n}\beta_{m}+\cdots\label{eq:}\end{equation}
where the quasiparticle operators $\beta_{m}$ and $\beta_{m}^{+}$
are the ones for which the operator $\hat{K}$ (and the matrix representation
$\mathbb{K}$) are diagonal and the product wave function $|\phi\rangle$
is the corresponding vacuum of the annihilation operators $\beta_{m}$.
We also notice in the previous expression of the statistical density
operator that the quantities $p_{m}$ can be interpreted as the {}``probability''
of a given quasiparticle excitation. By using the previous expression
of the statistical density operator we obtain \begin{equation}
\mathrm{Tr}[\hat{\rho}\hat{A}]=\langle\phi|\hat{A}|\phi\rangle+\sum_{m}p_{m}\langle\phi|\beta_{m}\hat{A}\beta_{m}^{+}|\phi\rangle+\frac{1}{2!}\sum_{mn}p_{m}p_{n}\langle\phi|\beta_{m}\beta_{n}\hat{A}\beta_{n}^{+}\beta_{m}^{+}|\phi\rangle+\cdots.\label{eq:}\end{equation}
that will prove to be useful in the following.

Let us now consider the generalization of Gaudin's theorem to the
case where the statistical density operator $\hat{\rho}$ is replaced
by the product $\hat{\rho}\hat{\mathcal{T}}^{-1}$ where $\hat{\mathcal{T}}$
is an operator performing a canonical transformation in the quasiparticle
operators $\alpha_{\mu}$ such that the transformed quasiparticle
operators\begin{equation}
\tilde{\alpha}_{\mu}=\hat{\mathcal{T}}\alpha_{\mu}\hat{\mathcal{T}}^{-1}=\sum_{\mu\nu}\mathbb{T}_{\mu\nu}\alpha_{\nu}\label{eq:}\end{equation}
still satisfy the canonical commutation relations but not necessarily
hermiticity. The requirement of preserving the canonical commutation
relations implies $\mathbb{T}\sigma\mathbb{T}^{T}=\sigma$. We now
have\begin{equation}
\alpha_{\mu}\hat{\rho}\hat{\mathcal{T}}^{-1}=\sum_{\nu}\rho_{\mu}\mathbb{T}_{\mu\nu}\hat{\rho}\hat{\mathcal{T}}^{-1}\alpha_{\nu}\label{eq:}\end{equation}
or $\tilde{\mathbb{M}}_{\mu\nu}=\rho_{\mu}\mathbb{T}_{\mu\nu}$ and
we are in the general situation that allowed us to derive Gaudin's
theorem. Therefore, for the evaluation of the trace\begin{equation}
\frac{\mathrm{Tr}[\hat{\rho}\hat{\mathcal{T}}^{-1}\hat{A]}}{\mathrm{Tr}[\hat{\rho}\hat{\mathcal{T}}^{-1}]}\label{eq:}\end{equation}
we can use Gaudin's theorem with the contractions given by Eq. (\ref{eq:contraction})
and with $\tilde{\mathbb{M}}=\rho_{\mu}\mathbb{T}_{\mu\nu}$ replacing
the matrix $\mathbb{M}$. Additionally, we have that\begin{equation}
\mathrm{Tr}[\hat{\rho}\hat{\mathcal{T}}^{-1}\hat{A}]=\langle\tilde{\phi}|\hat{A}|\phi\rangle+\sum_{m}p_{m}\langle\tilde{\phi}|\tilde{\beta}_{m}\hat{A}\beta_{m}^{+}|\phi\rangle+\frac{1}{2!}\sum_{mn}p_{m}p_{n}\langle\tilde{\phi}|\tilde{\beta}_{m}\tilde{\beta}_{n}\hat{A}\beta_{n}^{+}\beta_{m}^{+}|\phi\rangle+\cdots.\label{eq:TrDTA}\end{equation}
where $\langle\tilde{\phi}|=\langle\phi|\hat{\mathcal{T}}^{-1}$ and
$\tilde{\beta}_{m}=\hat{\mathcal{T}}\beta_{m}\hat{\mathcal{T}}^{-1}$.
In the limit $p_{m}\rightarrow0$ we have\begin{equation}
\frac{\langle\tilde{\phi}|\hat{A}|\phi\rangle}{\langle\tilde{\phi}|\phi\rangle}=\lim_{p_{m}\rightarrow0}\frac{\mathrm{Tr}[\hat{\rho}\hat{\mathcal{T}}^{-1}\hat{A}]}{\mathrm{Tr}[\hat{\rho}\hat{\mathcal{T}}^{-1}]}.\label{eq:}\end{equation}
This expression is the link between the results obtained with the
statistical density operator and the ones obtained with pure states
and it will allow to prove in the next section the extended Wick's
theorem out of the Gaudin's one.

\section{Proof of the extended wick's theorem}

In the previous section we have just shown that the overlap between
different mean field wave functions of a product of quasiparticle
operators can be written as the limit of the corresponding trace when
the statistical probabilities go to zero \begin{equation}
\frac{\langle\tilde{\phi}|\alpha_{\mu_{1}}\alpha_{\mu_{2}}\cdots\alpha_{\mu_{K}}|\phi\rangle}{\langle\tilde{\phi}|\phi\rangle}=\lim_{p_{m}\rightarrow0}\frac{\mathrm{Tr}[\hat{\rho}\hat{\mathcal{T}}^{-1}\alpha_{\mu_{1}}\alpha_{\mu_{2}}\cdots\alpha_{\mu_{K}}]}{\mathrm{Tr}[\hat{\rho}\hat{\mathcal{T}}^{-1}]}\label{eq:Limit0}\end{equation}
The argument of the limit of the right hand side of this equation
can be easily computed using Gaudin's theorem. It is given as the
sum of all possible contractions between pairs of quasiparticle operators
that are given by\begin{equation}
\mathbb{C}_{\mu_{1}\mu_{2}}=(1+\tilde{\mathbb{M}})_{\mu_{1}\overline{\mu}_{2}}^{-1}=[(1+\tilde{\mathbb{M}})^{-1}\sigma]_{\mu_{1}\mu_{2}}\label{eq:}\end{equation}
The relation given by Eq. (\ref{eq:Limit0}) tell us that the overlap
$\langle\tilde{\phi}|\alpha_{\mu_{1}}\alpha_{\mu_{2}}\cdots\alpha_{\mu_{K}}|\phi\rangle/\langle\tilde{\phi}|\phi\rangle$
is given as the product of all possible contractions $\overline{\mathbb{C}}_{\mu_{1}\mu_{2}}$
between pair of quasiparticle operators which are given by\begin{equation}
\overline{\mathbb{C}}_{\mu_{1}\mu_{2}}=\lim_{p_{m}\rightarrow0}\mathbb{C}{}_{\mu_{1}\mu_{2}}=\lim_{p_{m}\rightarrow0}[(1+\tilde{\mathbb{M}})^{-1}\sigma]_{\mu_{1}\mu_{2}}=\lim_{p_{m}\rightarrow0}[(1+\rho\mathbb{T})^{-1}\sigma]_{\mu_{1}\mu_{2}}\label{eq:}\end{equation}
provided that the limit exists. In order to compute the limit we have
to consider the general bipartite structure of the matrix $\mathbb{T}$\begin{equation}
\mathbb{T}=\left(\begin{array}{cc}
U & V\\
Y & X\end{array}\right)\label{eq:T-Bipartite}\end{equation}
in order to write \begin{eqnarray}
(1+\rho\mathbb{T})^{-1} & = & \left[\left(\begin{array}{cc}
1 & 0\\
0 & 1\end{array}\right)+\left(\begin{array}{cc}
p & 0\\
0 & p^{-1}\end{array}\right)\left(\begin{array}{cc}
U & V\\
Y & X\end{array}\right)\right]^{-1}\nonumber \\
 & = & \left[\left(\begin{array}{cc}
1 & 0\\
0 & p\end{array}\right)+\left(\begin{array}{cc}
p & 0\\
0 & 1\end{array}\right)\left(\begin{array}{cc}
U & V\\
Y & X\end{array}\right)\right]^{-1}\left(\begin{array}{cc}
1 & 0\\
0 & p\end{array}\right)\nonumber \\
 & = & \left[\left(\begin{array}{cc}
1 & 0\\
Y & X\end{array}\right)+\left(\begin{array}{cc}
p & 0\\
0 & p\end{array}\right)\left(\begin{array}{cc}
U & V\\
0 & 1\end{array}\right)\right]^{-1}\left(\begin{array}{cc}
1 & 0\\
0 & p\end{array}\right)\label{eq:Trick1}\end{eqnarray}
This expression allows a trivial evaluation of the $p_{\mu}\rightarrow0$
limit\begin{equation}
\lim_{p_{m}\rightarrow0}(1+\rho\mathbb{T})^{-1}=\left(\begin{array}{cc}
1 & 0\\
Y & X\end{array}\right)^{-1}\left(\begin{array}{cc}
1 & 0\\
0 & 0\end{array}\right)=\left(\begin{array}{cc}
1 & 0\\
-X^{-1}Y & 0\end{array}\right)\label{eq:limcontr}\end{equation}
that leads immediately to \begin{equation}
\overline{\mathbb{C}}_{\mu_{1}\mu_{2}}=\frac{\langle\tilde{\phi}|\alpha_{\mu_{1}}\alpha_{\mu_{2}}|\phi\rangle}{\langle\tilde{\phi}|\phi\rangle}=\left(\begin{array}{cc}
0 & 1\\
0 & -X^{-1}Y\end{array}\right)_{\mu_{1}\mu_{2}}\label{eq:}\end{equation}
or\begin{eqnarray}
\frac{\langle\tilde{\phi}|\beta_{m_{1}}\beta_{m_{2}}|\phi\rangle}{\langle\tilde{\phi}|\phi\rangle} & = & 0\label{eq:}\\
\frac{\langle\tilde{\phi}|\beta_{m_{1}}\beta_{m_{2}}^{+}|\phi\rangle}{\langle\tilde{\phi}|\phi\rangle} & = & \delta_{m_{1}m_{2}}\label{eq:}\\
\frac{\langle\tilde{\phi}|\beta_{m_{1}}^{+}\beta_{m_{2}}^{+}|\phi\rangle}{\langle\tilde{\phi}|\phi\rangle} & = & -(X^{-1}Y)_{m_{1}m_{2}}\label{eq:}\end{eqnarray}
which is the expected result of the GWT (see Refs. \cite{Balian.Brezin,Hara.79,Onishi.66}
for details). Another way to perform the limit is to use the property
of the matrix $\mathbb{T}$ (consequence of being the matrix of a
canonical transformation)\begin{equation}
\mathbb{T}\sigma\mathbb{T}^{T}\sigma=1\label{eq:TSTeqS}\end{equation}
in order to write $\mathbb{T}^{-1}=\sigma\mathbb{T}^{T}\sigma$. Using
this property it is very easy to show that \begin{equation}
(1+\rho\mathbb{T})^{-1}=\sigma\mathbb{T}^{T}\sigma(\sigma\mathbb{T}^{T}\sigma+\rho)^{-1}\label{eq:}\end{equation}
 This expression will prove to be useful in taking the limit as it
can be easily shown that \begin{equation}
\lim_{p_{m}\rightarrow0}\left[\left(\begin{array}{cc}
p & 0\\
0 & 1/p\end{array}\right)+\left(\begin{array}{cc}
A & B\\
C & D\end{array}\right)\right]^{-1}=\left(\begin{array}{cc}
A^{-1} & 0\\
0 & 0\end{array}\right).\label{eq:lim1}\end{equation}
Using now that\begin{equation}
\sigma\mathbb{T}^{T}\sigma=\left(\begin{array}{cc}
X^{T} & V^{T}\\
Y^{T} & U^{T}\end{array}\right)\label{eq:}\end{equation}
we easily arrive at\begin{equation}
\lim_{p_{m}\rightarrow0}\sigma\mathbb{T}^{T}\sigma(\sigma\mathbb{T}^{T}\sigma+\rho)^{-1}=\left(\begin{array}{cc}
X^{T} & V^{T}\\
Y^{T} & U^{T}\end{array}\right)\left(\begin{array}{cc}
X^{T-1} & 0\\
0 & 0\end{array}\right)=\left(\begin{array}{cc}
1 & 0\\
Y^{T}X^{T-1} & 0\end{array}\right)\label{eq:}\end{equation}
which is equivalent to Eq. (\ref{eq:limcontr}) because $Y^{T}X^{T-1}=-X^{-1}Y$
as can be deduced from the property of Eq. (\ref{eq:TSTeqS}). 

As a demonstration of the usefulness of the method just described
we will consider the generalized Wick's theorem for the quantity\begin{equation}
\frac{\langle\tilde{\phi}_{1}|\alpha_{\mu_{1}}\alpha_{\mu_{2}}\cdots\alpha_{\mu_{K}}|\tilde{\phi}_{2}\rangle}{\langle\tilde{\phi}_{1}|\tilde{\phi}_{2}\rangle}\label{eq:}\end{equation}
where the quasiparticle operators $\alpha_{\mu_{k}}$ are not related
to the mean field wave functions $|\tilde{\phi}_{i}\rangle$ of the
previous overlap. The above overlap can be written as the following
limit\begin{equation}
\frac{\langle\tilde{\phi}_{1}|\alpha_{\mu_{1}}\alpha_{\mu_{2}}\cdots\alpha_{\mu_{K}}|\tilde{\phi}_{2}\rangle}{\langle\tilde{\phi}_{1}|\tilde{\phi}_{2}\rangle}=\lim_{p_{m}\rightarrow0}\frac{\mathrm{Tr}[\hat{\rho}\hat{\mathcal{T}}_{1}^{-1}\alpha_{\mu_{1}}\alpha_{\mu_{2}}\cdots\alpha_{\mu_{K}}\hat{\mathcal{T}}_{2}]}{\mathrm{Tr}[\hat{\rho}\hat{\mathcal{T}}_{1}^{-1}\hat{\mathcal{T}}_{2}]}\label{eq:}\end{equation}
where it has been assumed that there exist operators $\hat{\mathcal{T}}_{i}$
relating the mean field wave functions $|\tilde{\phi}_{i}\rangle$
with the vacuum of the quasiparticle annihilation operators of the
generalized set $\alpha_{\mu}$ ($\beta_{k}|\phi\rangle=0)$ through
the relation $|\tilde{\phi}_{i}\rangle=\hat{\mathcal{T}}_{i}|\phi\rangle$.
Considering the extended Gaudin's theorem of the previous section
we can write the argument of the limit of the right hand side as the
sum of all possible contractions\begin{equation}
\mathbb{C}_{\mu_{1}\mu_{2}}=(1+\mathbb{T}_{2}^{-1}\rho\mathbb{T}_{1})_{\mu_{1}\overline{\mu}_{2}}^{-1}=\left(\mathbb{T}_{1}^{-1}(\mathbb{T}_{2}\mathbb{T}_{1}^{-1}+\rho)^{-1}\mathbb{T}_{2}\right)_{\mu_{1}\overline{\mu}_{2}}=\left(\sigma\mathbb{T}_{1}^{T}\sigma(\mathbb{T}_{2}\sigma\mathbb{T}_{1}^{T}\sigma+\rho)^{-1}\mathbb{T}_{2}\sigma\right)_{\mu_{1}\mu_{2}}\label{eq:}\end{equation}
The limit $p_{m}\rightarrow0$ in this expression can be evaluated
straightforwardly by using the result of Eq. (\ref{eq:lim1}) and
the notation of Eq. (\ref{eq:T-Bipartite})\begin{equation}
\overline{\mathbb{C}}=\lim_{p_{m}\rightarrow0}\mathbb{C}=\left(\begin{array}{cc}
X_{1}^{T} & V_{1}^{T}\\
Y_{1}^{T} & U_{1}^{T}\end{array}\right)\left(\begin{array}{cc}
(U_{2}X_{1}^{T}+V_{2}Y_{1}^{T})^{-1} & 0\\
0 & 0\end{array}\right)\left(\begin{array}{cc}
V_{2} & U_{2}\\
X_{2} & Y_{2}\end{array}\right)\label{eq:Contract2}\end{equation}
Finally, we shall derive all the contractions needed for the evaluation
of the most general overlap

\begin{equation}
\frac{\langle\tilde{\phi}_{1}|\alpha_{\mu_{1}}\alpha_{\mu_{2}}\cdots\alpha_{\mu_{j}}\hat{\mathcal{T}}_{3}\alpha_{\mu_{j+1}}\cdots\alpha_{\mu_{K}}|\tilde{\phi}_{2}\rangle}{\langle\tilde{\phi}_{1}|\hat{\mathcal{T}}_{3}|\tilde{\phi}_{2}\rangle}=\lim_{p_{m}\rightarrow0}\frac{\mathrm{Tr}[\hat{\rho}\hat{\mathcal{T}}_{1}^{-1}\alpha_{\mu_{1}}\alpha_{\mu_{2}}\cdots\alpha_{\mu_{j}}\hat{\mathcal{T}}_{3}\alpha_{\mu_{j+1}}\cdots\alpha_{\mu_{K}}\hat{\mathcal{T}}_{2}]}{\mathrm{Tr}[\hat{\rho}\hat{\mathcal{T}}_{1}^{-1}\hat{\mathcal{T}}_{3}\hat{\mathcal{T}}_{2}]}.\label{eq:}\end{equation}
In this case, the operator $\hat{\mathcal{T}}_{3}$ separates the
product of quasiparticle operators in two groups, the ones to the
left of this operator and the ones to the right. Because of this separation
we will need to consider three different kinds of contractions depending
upon in which group are located each of the two quasiparticle operators
involved in the contraction. To be more specific, we will need the
contractions\begin{equation}
\mathbb{C}_{\mu\nu}^{(1)}=\left((1+\mathbb{T}_{3}^{-1}\mathbb{T}_{2}^{-1}\rho\mathbb{T}_{1})^{-1}\sigma\right)_{\mu\nu}\label{eq:Contract31}\end{equation}
when both indices $\mu$ and $\nu$ are in the set of operators to
the left of $\mathcal{T}_{3}$ and with indices in the set $\{\mu_{1},\ldots,\mu_{j}\}$,
\begin{equation}
\mathbb{C}_{\mu\nu}^{(2)}=\left((1+\mathbb{T}_{3}^{-1}\mathbb{T}_{2}^{-1}\rho\mathbb{T}_{1})^{-1}\mathbb{T}_{3}\sigma\right)_{\mu\nu}\label{eq:Contract32}\end{equation}
when $\mu$ is in the set $\{\mu_{1},\ldots,\mu_{j}\}$ (i.e. to the
left) and $\nu$ is in the set $\{\mu_{j+1},\ldots,\mu_{K}\}$ (i.e.
to the right) and finally \begin{equation}
\mathbb{C}_{\mu\nu}^{(3)}=\left((1+\mathbb{T}_{2}^{-1}\rho\mathbb{T}_{1}\mathbb{T}_{3}^{-1})^{-1}\sigma\right)_{\mu\nu}\label{eq:Contract33}\end{equation}
when both indices $\mu$ and $\nu$ belong to quasiparticle operators
that are to the right of $\hat{\mathcal{T}}_{3}$ and therefore are
in the set $\{\mu_{j+1},\ldots,\mu_{K}\}$. The limits $p_{m}\rightarrow0$
in the above contractions can be very easily performed using previous
considerations like the ones leading to Eq. (\ref{eq:Contract2})
and will not be given here.

\section{Evaluation of the partition function or norm overlap}

In this section we will evaluate the expression for the partition
function or norm overlap appearing in the previous section. A typical
partition function to evaluate is of the form $\mathrm{Tr}[\hat{\rho}\hat{\mathcal{T}}_{1}^{-1}]$.
Taking into account that the density operator also performs a canonical
transformation, the product $\hat{\rho}\hat{\mathcal{T}}_{1}^{-1}$
also corresponds to a canonical transformation whose matrix representation
is the product of the matrix representation of $\hat{\rho}$ and $\mathcal{T}_{1}^{-1}$.
Therefore, we could reduce the calculation of $\mathrm{Tr}[\hat{\rho}\hat{\mathcal{T}}_{1}^{-1}]$
to the one of $\mathrm{Tr}[\hat{\mathcal{T}}]$ where $\hat{\rho}\hat{\mathcal{T}}_{1}^{-1}=\hat{\mathcal{T}}$.
However, in the calculations of the traces we will have to assume
that at least one of the canonical transformations can be written
as the exponential of an one body operator and therefore the evaluation
of $\mathrm{Tr}[\hat{\rho}\hat{\mathcal{T}}_{1}^{-1}]$ can be considered
as more general by assuming that the transformation $\hat{\mathcal{T}}_{1}^{-1}$
is given by the exponential of an one body operator, namely $\hat{\mathcal{T}}_{1}^{-1}=e^{-\hat{S}_{1}}=\exp(-\frac{1}{2}\sum_{\mu\nu}\alpha_{\mu}\left(\mathbb{S}_{1}\right)_{\mu\nu}\alpha_{\nu})$
and considering that $\hat{\rho}$ is a general operator performing
a canonical transformation and not necessarily expressible as the
exponential of an one body operator (see appendix A). To compute $\mathrm{Tr}[\hat{\rho}\hat{\mathcal{T}}_{1}^{-1}]$
we consider the function\begin{equation}
n(\lambda)=\mathrm{Tr}[\hat{\rho}\exp(-\lambda\hat{S}_{1})]\label{eq:}\end{equation}
that reduces to the quantity we want to compute in the limit $\lambda=1$.
According to previous notation the operator $\hat{S}_{1}$ is given
by $\hat{S}_{1}=\frac{1}{2}\sum_{\mu\nu}\alpha_{\mu}\left(\mathbb{S}_{1}\right)_{\mu\nu}\alpha_{\nu}$.
To evaluate $n(\lambda)$ we consider its derivative with respect
to $\lambda$ \begin{equation}
\frac{dn(\lambda)}{d\lambda}=-\mathrm{Tr}[\hat{\rho}\exp(-\lambda\hat{S}_{1})\hat{S}_{1}]\label{eq:}\end{equation}
that can be easily computed with the help of the extended Gaudin's
theorem considered in the previous section \begin{equation}
\frac{dn(\lambda)}{d\lambda}=-\frac{1}{2}\sum_{\mu\nu}\left(\mathbb{S}_{1}\right)_{\mu\nu}\mathrm{Tr}[\hat{\rho}\exp(-\lambda\hat{S}_{1})\alpha_{\mu}\alpha_{\nu}]=-\frac{1}{2}\sum_{\mu\nu}\left(\mathbb{S}_{1}\right)_{\mu\nu}\left((1+\mathbb{M}\mathbb{T}(\lambda))^{-1}\sigma\right)_{\mu\nu}n(\lambda)\label{eq:}\end{equation}
where, as usual, the transformation matrix $\mathbb{T}(\lambda)$
is given by \begin{equation}
\mathbb{T}(\lambda)=\exp(-\lambda\sigma\mathbb{S}_{1,A})\label{eq:}\end{equation}
with $\mathbb{S}_{1,A}=\frac{1}{2}\left(\mathbb{S}_{1}-\mathbb{S}_{1}^{T}\right)$
and $\mathbb{M}$ is given by Eq. (\ref{eq:aro}). Using the property
$(1+\mathbb{M}\mathbb{T}(\lambda))^{-1}\sigma+\sigma(1+\mathbb{T}^{T}(\lambda)\mathbb{M}^{T})^{-1}=\sigma$
deduced in appendix A and a little of algebra we obtain \begin{equation}
\frac{1}{n(\lambda)}\frac{dn(\lambda)}{d\lambda}=\frac{1}{2}\mathrm{Tr}[\sigma\mathbb{S}_{1,A}(1+\mathbb{M}\mathbb{T}(\lambda))^{-1}]-\frac{1}{2}\mathrm{Tr}[\sigma\mathbb{S}_{1}]\label{eq:}\end{equation}
The first term of the right hand side of the equation can be written
(see appendix B) as\begin{equation}
\mathrm{Tr}[\sigma\mathbb{S}_{1,A}(1+\mathbb{M}\mathbb{T}(\lambda))^{-1}]=\frac{d}{d\lambda}\mathrm{Tr}[\ln(1+\mathbb{M}\mathbb{T}(\lambda)]\label{eq:}\end{equation}
which allows to integrate the differential equation defining $n(\lambda)$
to obtain \begin{equation}
n(\lambda)=Ce^{-\frac{\lambda}{2}\mathrm{Tr}[\sigma\mathbb{S}_{1}]}[\det(1+\mathbb{M}\mathbb{T}(\lambda))]^{1/2}\label{eq:}\end{equation}
where $C$ is an arbitrary constant and the relation $\det A=\exp(\mathrm{Tr}[\ln A])$
has been used. The constant $C$ is determined by taking into account
that $n(0)=\mathrm{Tr}[\hat{\rho}]=C[\det(1+\mathbb{M})]^{1/2}$ so
that we finally arrive to the general expression \begin{equation}
\frac{\mathrm{Tr}[\hat{\rho}\hat{\mathcal{T}}_{1}^{-1}]}{\mathrm{Tr}[\hat{\rho}]}=e^{-\frac{1}{2}\mathrm{Tr}[\sigma\mathbb{S}_{1}]}\frac{[\det(1+\mathbb{M}\mathbb{T}(1))]^{1/2}}{[\det(1+\mathbb{M})]^{1/2}}\label{eq:}\end{equation}
In the very common case where $\hat{\rho}$ is also given as the exponential
of an one body operator we can choose $\hat{\rho}=\hat{\mathcal{T}}_{1}$
and deduce that \begin{equation}
\mathrm{Tr}[\hat{\rho}]=e^{\frac{1}{2}\mathrm{Tr}[\sigma\mathbb{S}_{\rho}]}[\det(1+\mathbb{M})]^{1/2}\label{eq:}\end{equation}
where we have made use of the property $\mathbb{M}\mathbb{T}(1)=\openone$
holding in this specific case. Taking everything into account we arrive
at the particular result\begin{equation}
\mathrm{Tr}[\hat{\rho}\hat{\mathcal{T}}_{1}^{-1}]=e^{-\frac{1}{2}\mathrm{Tr}[\sigma(\mathbb{S}_{1}-\mathbb{S}_{\rho})]}[\det(1+\mathbb{M}\mathbb{T}(1))]^{1/2}.\label{eq:TrThoT}\end{equation}

In the case where $\hat{\rho}$ is {}``diagonal'' in the sense that
it is given by \begin{equation}
\hat{\rho}=\exp(\sum_{m}k_{m}\beta_{m}^{+}\beta_{m})=\exp(\frac{1}{2}\sum_{m}k_{m})\exp(\frac{1}{2}\sum_{\mu\nu}\alpha_{\mu}\left(\mathbb{S}_{\rho}\right)_{\mu\nu}\alpha_{\nu})\label{eq:Rho-diag}\end{equation}
with\begin{equation}
\mathbb{S}_{\rho}=\left(\begin{array}{cc}
0 & -k\\
k & 0\end{array}\right)\label{eq:}\end{equation}
the transformation matrix $\mathbb{M}$ is diagonal and given in terms
of the probabilities $p_{k}$ and their inverses\begin{equation}
\mathbb{M}=\exp(\sigma\mathbb{S}_{\rho})=\left(\begin{array}{cc}
p & 0\\
0 & p^{-1}\end{array}\right)\equiv\rho.\label{eq:}\end{equation}
The constant term in Eq. ( \ref{eq:Rho-diag}) is written in terms
of the probabilities as \begin{equation}
\exp(\frac{1}{2}\sum_{m}k_{m})=\left(\prod_{m}p_{m}\right)^{1/2}=\left[\det\left(\begin{array}{cc}
1 & 0\\
0 & p\end{array}\right)\right]^{1/2}\label{eq:}\end{equation}
Using all these definitions we can finally write \begin{equation}
\hat{\rho}\hat{\mathcal{T}}_{1}^{-1}=\left(\prod_{m}p_{m}\right)^{1/2}\exp(\frac{1}{2}\sum_{\mu\nu}\alpha_{\mu}\left(\mathbb{S}_{\rho}\right)_{\mu\nu}\alpha_{\nu})\exp(-\frac{1}{2}\sum_{\mu\nu}\alpha_{\mu}\left(\mathbb{S}_{1}\right)_{\mu\nu}\alpha_{\nu})\label{eq:}\end{equation}
 that leads, by means of Eq. (\ref{eq:TrThoT}), to\begin{equation}
\mathrm{Tr}[\hat{\rho}\hat{\mathcal{T}}_{1}^{-1}]=\left(\prod_{m}p_{m}\right)^{1/2}e^{-\frac{1}{2}\mathrm{Tr}[\sigma\mathbb{S}_{1}]}[\det(1+\rho\mathbb{T}_{1})]^{1/2}=e^{-\frac{1}{2}\mathrm{Tr}[\sigma\mathbb{S}_{1}]}\left[\det\left(\begin{array}{cc}
1 & 0\\
0 & p\end{array}\right)\right]^{1/2}[\det(1+\rho\mathbb{T}_{1})]^{1/2}\label{eq:}\end{equation}
which is finally written as \begin{equation}
\mathrm{Tr}[\hat{\rho}\hat{\mathcal{T}}_{1}^{-1}]=e^{-\frac{1}{2}\mathrm{Tr}[\sigma\mathbb{S}_{1}]}\left[\det\left(\left(\begin{array}{cc}
1 & 0\\
0 & p\end{array}\right)+\left(\begin{array}{cc}
p & 0\\
0 & 1\end{array}\right)\mathbb{T}_{1}\right)\right]^{1/2}\label{eq:}\end{equation}
Now the $p\rightarrow0$ limit is straightforward\begin{equation}
\langle\tilde{\phi}_{1}|\phi\rangle=\lim_{p_{m}\rightarrow0}\mathrm{Tr}[\hat{\rho}\hat{\mathcal{T}}_{1}^{-1}]=e^{-\frac{1}{2}\mathrm{Tr}[\sigma\mathbb{S}_{1}]}\left[\det\left(\left(\begin{array}{cc}
1 & 0\\
Y_{1} & X_{1}\end{array}\right)\right)\right]^{1/2}=e^{-\frac{1}{2}\mathrm{Tr}[\sigma\mathbb{S}_{1}]}\left[\det X_{1}\right]^{1/2}\label{eq:}\end{equation}
and corresponds to the expected result \cite{Onishi.66,Balian.Brezin,Hara.79}.

\section{Multiquasiparticle overlaps}

Finally, we will consider a very helpful extension of the above considerations
to the evaluation of overlaps between multiquasiparticle excitations
\begin{equation}
\frac{\left\langle \tilde{\phi}\right|\tilde{\beta}_{\mu_{1}}\tilde{\beta}_{\mu_{2}}\cdots\tilde{\beta}_{\mu_{K}}\hat{A}\bar{\beta}_{\mu_{K}}\cdots\bar{\beta}_{\mu_{2}}\bar{\beta}_{\mu_{1}}\left|\phi\right\rangle }{\left\langle \tilde{\phi}\right|\tilde{\beta}_{\mu_{1}}\tilde{\beta}_{\mu_{2}}\cdots\tilde{\beta}_{\mu_{K}}\bar{\beta}_{\mu_{K}}\cdots\bar{\beta}_{\mu_{2}}\bar{\beta}_{\mu_{1}}\left|\phi\right\rangle }=\frac{\left\langle \phi\right|\beta_{\mu_{1}}\beta_{\mu_{2}}\cdots\beta_{\mu_{K}}\hat{\mathcal{T}}^{-1}\hat{A}\bar{\beta}_{\mu_{K}}\cdots\bar{\beta}_{\mu_{2}}\bar{\beta}_{\mu_{1}}\left|\phi\right\rangle }{\left\langle \phi\right|\beta_{\mu_{1}}\beta_{\mu_{2}}\cdots\beta_{\mu_{K}}\hat{\mathcal{T}}^{-1}\bar{\beta}_{\mu_{K}}\cdots\bar{\beta}_{\mu_{2}}\bar{\beta}_{\mu_{1}}\left|\phi\right\rangle }\label{eq:Multiqp-over}\end{equation}
where the {}``tilded'' quasiparticle operators are related to the
{}``untilded'' ones through a canonical transformation\begin{equation}
\tilde{\alpha}_{\mu}=\hat{\mathcal{T}}\alpha_{\mu}\hat{\mathcal{T}}^{-1}=\sum_{\nu}\mathbb{T}_{\mu\nu}\alpha_{\nu}\label{eq:}\end{equation}
and $|\tilde{\phi}\rangle=\hat{\mathcal{T}}|\phi\rangle$. In the
spirit of the preceding sections the above overlap can be written
as \begin{equation}
\frac{\left\langle \phi\right|\beta_{\mu_{1}}\beta_{\mu_{2}}\cdots\beta_{\mu_{K}}\hat{\mathcal{T}}^{-1}\hat{A}\bar{\beta}_{\mu_{K}}\cdots\bar{\beta}_{\mu_{2}}\bar{\beta}_{\mu_{1}}\left|\phi\right\rangle }{\left\langle \phi\right|\beta_{\mu_{1}}\beta_{\mu_{2}}\cdots\beta_{\mu_{K}}\hat{\mathcal{T}}^{-1}\bar{\beta}_{\mu_{K}}\cdots\bar{\beta}_{\mu_{2}}\bar{\beta}_{\mu_{1}}\left|\phi\right\rangle }=\lim_{[p\rightarrow\infty]}\frac{\mathrm{Tr}[\hat{\rho}\hat{\mathcal{T}}^{-1}\hat{A}]}{\mathrm{Tr}[\hat{\rho}\hat{\mathcal{T}}^{-1}]}\label{eq:Quasip-overlap}\end{equation}
where the limit has to be understood as a limiting process where the
probabilities $p_{\mu_{1}},\ldots,p_{\mu_{K}}$ are made to tend to
infinity while at the same time the other probabilities $p_{\nu}$
($\nu\ne\mu_{1},\ldots,\mu_{K}$) are set to zero. In fact, for the
present purposes there is no need to consider such a general limit
and it is enough to consider that all the $p_{\mu_{1}},\ldots,p_{\mu_{K}}$
that tend to infinity do so in the same way, that is $p_{\mu_{1}}=p_{\mu_{2}}=\ldots=p_{\mu_{K}}=p$.
Therefore, in the numerator of the right hand side of Eq. (\ref{eq:Quasip-overlap})
the greatest power of $p$ ($p^{K}$) will correspond to the numerator
of the left hand side of the same equation. The same holds true for
the denominators and therefore taking the limit $p\rightarrow\infty$
will give the desired result. By using the generalized Gaudin's theorem
we can write the trace of Eq. (\ref{eq:Quasip-overlap}) in terms
of the contractions\begin{equation}
\mathbb{C}_{\rho\sigma}=\left[\left(1+\mathbb{\rho T}\right)^{-1}\sigma\right]_{\rho\sigma}\label{eq:}\end{equation}
and therefore in order to evaluate the left hand side of Eq. (\ref{eq:Quasip-overlap})
we will have to consider the matrix of contractions \begin{equation}
\overline{\mathbb{C}}_{\rho\sigma}=\lim_{[p\rightarrow\infty]}\mathbb{C}_{\rho\sigma}=\lim_{[p\rightarrow\infty]}\left[\left(1+\mathbb{\rho T}\right)^{-1}\sigma\right]_{\rho\sigma}\label{eq:1}\end{equation}
To obtain the explicit expression for the above limit we will take
into account the bipartite structure of the matrix $\mathbb{T}$ given
in Eq. (\ref{eq:T-Bipartite}) as well as Eqs. (\ref{eq:Trick1})
and (\ref{eq:CanonicalCond}) to write\begin{equation}
\mathbb{C}=\left(\begin{array}{cc}
1 & 0\\
Y^{T}X^{T-1} & X^{-1}\end{array}\right)\left[\mathbb{I}+\left(\begin{array}{cc}
p & 0\\
0 & p\end{array}\right)\left(\begin{array}{cc}
X^{T-1} & {VX}^{-1}\\
Y^{T}X^{T-1} & X^{-1}\end{array}\right)\right]^{-1}\left(\begin{array}{cc}
1 & 0\\
0 & p\end{array}\right)\sigma\label{eq:MQP1}\end{equation}
The limit $[p\rightarrow\infty]$ will be taken in two steps, first
the probabilities $p_{\nu}$ with $\nu\ne\mu_{1},\ldots,\mu_{K}$
will be set to zero and the remaining $p_{\mu_{1}},p_{\mu_{2}}$,
etc will be taken as equal to an unique parameter $p$ that will be
made to tend to infinity afterwards. By applying the first step (i.e.
setting to zero the $p_{\nu}$ ($\nu\ne\mu_{1},\ldots,\mu_{K}$))
the argument of the inverse matrix of Eq. (\ref{eq:MQP1}) will be
the sum of the identity matrix plus a matrix where the only surviving
rows will correspond to those of the probabilities $p_{\mu_{j}}$
going to infinity, that is\begin{equation}
\mathbb{I}+\left(\begin{array}{cc}
p & 0\\
0 & p\end{array}\right)\left(\begin{array}{cc}
X^{T-1} & {VX}^{-1}\\
Y^{T}X^{T-1} & X^{-1}\end{array}\right)=\mathbb{I}+p\left(\begin{array}{cccccccc}
0 & 0 & \cdots & 0 & 0 & 0 & \cdots & 0\\
\vdots & \vdots & \cdots & \vdots & \vdots & \vdots & \cdots & \vdots\\
R_{\mu_{1}1} & R_{\mu_{1}2} & \cdots & R_{\mu_{1}N} & T_{\mu_{1}1} & T_{\mu_{1}2} & \cdots & T_{\mu_{1}N}\\
\vdots & \vdots & \cdots & \vdots & \vdots & \vdots & \cdots & \vdots\\
R_{\mu_{K}1} & R_{\mu_{K}2} & \cdots & R_{\mu_{K}N} & T_{\mu_{K}1} & T_{\mu_{K}2} & \cdots & T_{\mu_{K}N}\\
\vdots & \vdots & \cdots & \vdots & \vdots & \vdots & \cdots & \vdots\\
0 & 0 & \cdots & 0 & 0 & 0 & \cdots & 0\\
\vdots & \vdots & \cdots & \vdots & \vdots & \vdots & \cdots & \vdots\\
S_{\mu_{1}1} & S_{\mu_{1}2} & \cdots & S_{\mu_{1}N} & W_{\mu_{1}1} & W_{\mu_{1}2} & \cdots & W_{\mu_{1}N}\\
\vdots & \vdots & \cdots & \vdots & \vdots & \vdots & \cdots & \vdots\\
S_{\mu_{K}1} & S_{\mu_{K}2} & \cdots & S_{\mu_{K}N} & W_{\mu_{K}1} & W_{\mu_{K}2} & \cdots & W_{\mu_{k}N}\\
\vdots & \vdots & \cdots & \vdots & \vdots & \vdots & \cdots & \vdots\\
0 & 0 & \cdots & 0 & 0 & 0 & \cdots & 0\end{array}\right)\label{eq:MQP-LIM1}\end{equation}
where we have set $R=X^{T-1}$, $T=VX^{-1}$, $S=Y^{T}X^{T-1}$ and
$W=X^{-1}$ to lighten a little bit the notation. Now it is convenient
to introduce the vectors\begin{equation}
r_{\mu_{j}}=\left(\begin{array}{c}
R_{\mu_{j}1}\\
\vdots\\
R_{\mu_{j}N}\\
T_{\mu_{j}1}\\
\vdots\\
T_{\mu_{j}N}\end{array}\right),\: s_{\mu_{j}}=\left(\begin{array}{c}
S_{\mu_{j}1}\\
\vdots\\
S_{\mu_{j}N}\\
W_{\mu_{j}1}\\
\vdots\\
W_{\mu_{j}N}\end{array}\right)\label{eq:}\end{equation}
and denote by $e_{\mu_{j}}$ the Cartesian basis vector of dimension
$2N$ along the direction $\mu_{j}$. The introduction of these quantities
allow us to write Eq. (\ref{eq:MQP-LIM1}) in a more tractable and
compact form as\begin{equation}
\mathbb{I}+p\sum_{j=1}^{K}e_{\mu_{j}}\otimes r_{\mu_{j}}+e_{\mu_{j}+N}\otimes s_{\mu_{j}}\label{eq:}\end{equation}
The expression of this matrix can be further simplified by introducing
the $2N\times2K$ matrices $E=(e_{\mu_{1}}\cdots e_{\mu_{K}}e_{\mu_{1}+N}\cdots e_{\mu_{K}+N})$
and $Q=(r_{\mu_{1}}\cdots r_{\mu_{K}}s_{\mu_{1}}\cdots s_{\mu_{K}})$
in order to obtain the final result\begin{equation}
\openone+\left(\begin{array}{cc}
p & 0\\
0 & p\end{array}\right)\left(\begin{array}{cc}
X^{T-1} & {VX}^{-1}\\
Y^{T}X^{T-1} & X^{-1}\end{array}\right)=\openone+pEQ^{T}\label{eq:}\end{equation}
The inverse matrix can be computed with the help of the Woodbury formula
\cite{VanLoan.96} \begin{equation}
\left(\mathbb{I}+pEQ^{T}\right)^{-1}=\mathbb{I}-pE\left(\mathbb{I}+pQ^{T}E\right)^{-1}Q^{T}\label{eq:}\end{equation}
where now the matrix to be inverted $\left(\mathbb{I}+pQ^{T}E\right)$
is a matrix of dimension $2K\times2K$ instead of $2N\times2N$. With
the $p\rightarrow\infty$ limit in mind we will expand the right hand
side of the previous equation in a power series of the inverse of
$p$ as\begin{equation}
\left(\mathbb{I}+pEQ^{T}\right)^{-1}=\mathbb{I}-E(Q^{T}E)^{-1}Q^{T}+\frac{1}{p}E(Q^{T}E)^{-2}Q^{T}+\cdots\label{eq:}\end{equation}
In a first sight one could think that this quantity, when multiplied
by the remaining matrix depending upon $p$, will lead to a divergent
quantity after the $p\rightarrow\infty$ limit has been taken. Fortunately,
this is not the case and in order to recognize how different terms
cancel out it is convenient to use the following identity \begin{equation}
\left(\begin{array}{cc}
1 & 0\\
0 & p\end{array}\right)=\left(\begin{array}{cc}
1 & 0\\
0 & 0\end{array}\right)+pEE^{T}\left(\begin{array}{cc}
0 & 0\\
0 & 1\end{array}\right)\label{eq:}\end{equation}
that is only valid after the limit $p_{\nu}\rightarrow0$ ($\nu\ne\mu_{1},\ldots,\mu_{k}$)
has been taken. The product $\left(\mathbb{I}+pEQ^{T}\right)^{-1}\left(\begin{array}{cc}
1 & 0\\
0 & p\end{array}\right)$ now simplifies owing to the fact that $\left(\mathbb{I}-E(Q^{T}E)^{-1}Q^{T}\right)EE^{T}=EE^{T}-EE^{T}=0$.
Combining all the previous results together leads to the final result
(with the limit $p\rightarrow\infty$) already taken\begin{equation}
\mathbb{\overline{C}}=\left(\begin{array}{cc}
1 & 0\\
Y^{T}X^{T-1} & X^{-1}\end{array}\right)\left[\left(\begin{array}{cc}
1 & 0\\
0 & 0\end{array}\right)+E(Q^{T}E)^{-1}\tilde{Q}^{T}\right]\sigma\label{eq:Multiqp-contr}\end{equation}
with $\tilde{Q}^{T}=E^{T}\left(\begin{array}{cc}
0 & 0\\
0 & 1\end{array}\right)-Q^{T}\left(\begin{array}{cc}
1 & 0\\
0 & 0\end{array}\right)$. This result is quite relevant as it synthesizes in just one very
compact formula the combinatorial number of contractions needed for
the evaluation of the multiquasiparticle overlap of Eq. (\ref{eq:Multiqp-contr}).
The evaluation of the multiquasiparticle overlap could of course be
carried out by means of the standard GWT but it would be a very painful
procedure as the number of terms to be considered increases dramatically
(combinatorially) with the number of multiquasiparticle excitations
(for instance, if we consider a four quasiparticle excitation overlap
of a two body operator $\hat{A}$ (this is not such an uncommon overlap,
see \cite{Hara.95} for more details) then we should consider a twelve
quasiparticle matrix element that involves $11!!$ (that is 10395)
contractions instead of just the four need by using Eq. (\ref{eq:Multiqp-contr}).
The above formula also presents computational advantages if one has
to consider a variety of multiquasiparticle excitations. The reason
is that it allows to split the calculation of the contractions in
two well differentiated steps; the first is common to all the multiquasiparticle
excitations to be considered and consist of the evaluation of the
first matrix of Eq. (\ref{eq:Multiqp-contr}). This is the most expensive
computation from a computational point of view as it involves the
inversion of the matrix $X$ that is of dimension $N\times N$. The
second step depends on the indices of the multiquasiparticle excitations
considered and consists of the construction of the matrices $E$ and
$Q$ (of dimension $2N\times2K$) out of the rows of the corresponding
matrices and the inversion of the $2K\times2K$ matrix $Q^{T}E$.
For instance, in the case of a four quasiparticle excitation $K=4$
and the later matrix to be inverted is of dimension $8\times8$. At
this point, we can ask whether it is possible to generalize the multiquasiparticle
overlap of Eq. (\ref{eq:Quasip-overlap}) to the most general situation
where the indices on the right multiquasiparticle excitation differ
from the ones of the excitation acting on the left, namely\begin{equation}
\frac{\langle\tilde{\phi}|\tilde{\beta}_{\nu_{1}}\tilde{\beta}_{\nu_{2}}\cdots\tilde{\beta}_{\nu_{K}}\hat{A}\bar{\beta}_{\mu_{K}}\cdots\bar{\beta}_{\mu_{2}}\bar{\beta}_{\mu_{1}}\left|\phi\right\rangle }{\langle\tilde{\phi}|\tilde{\beta}_{\nu_{1}}\tilde{\beta}_{\nu_{2}}\cdots\tilde{\beta}_{\nu_{K}}\bar{\beta}_{\mu_{K}}\cdots\bar{\beta}_{\mu_{2}}\bar{\beta}_{\mu_{1}}\left|\phi\right\rangle }\label{eq:Multiqp-overmunu}\end{equation}
The answer is affirmative and involves the introduction of an operator
$\hat{\mathcal{T}}\left(\begin{array}{c}
\mu\\
\nu\end{array}\right)$ that transforms the indices $\nu$ into the indices $\mu$ in the
following way\begin{equation}
\left\langle \phi\right|\beta_{\mu_{1}}\beta_{\mu_{2}}\cdots\beta_{\mu_{K}}\hat{\mathcal{T}}\left(\begin{array}{c}
\mu\\
\nu\end{array}\right)=\left\langle \phi\right|\beta_{\nu_{1}}\beta_{\nu_{2}}\cdots\beta_{\nu_{K}}.\label{eq:}\end{equation}
By introducing this transformation operator we can express the overlap
of Eq. (\ref{eq:Multiqp-overmunu}) in the more familiar form of Eq.
(\ref{eq:Multiqp-over}) \begin{equation}
\frac{\left\langle \tilde{\phi}\right|\tilde{\beta}_{\nu_{1}}\tilde{\beta}_{\nu_{2}}\cdots\tilde{\beta}_{\nu_{K}}\hat{A}\bar{\beta}_{\mu_{K}}\cdots\bar{\beta}_{\mu_{2}}\bar{\beta}_{\mu_{1}}\left|\phi\right\rangle }{\left\langle \tilde{\phi}\right|\tilde{\beta}_{\nu_{1}}\tilde{\beta}_{\nu_{2}}\cdots\tilde{\beta}_{\nu_{K}}\bar{\beta}_{\mu_{K}}\cdots\bar{\beta}_{\mu_{2}}\bar{\beta}_{\mu_{1}}\left|\phi\right\rangle }=\frac{\left\langle \phi\right|\beta_{\mu_{1}}\beta_{\mu_{2}}\cdots\beta_{\mu_{K}}\hat{\mathcal{T}}\left(\begin{array}{c}
\mu\\
\nu\end{array}\right)\hat{\mathcal{T}}^{-1}\hat{A}\bar{\beta}_{\mu_{K}}\cdots\bar{\beta}_{\mu_{2}}\bar{\beta}_{\mu_{1}}\left|\phi\right\rangle }{\left\langle \phi\right|\beta_{\mu_{1}}\beta_{\mu_{2}}\cdots\beta_{\mu_{K}}\hat{\mathcal{T}}\left(\begin{array}{c}
\mu\\
\nu\end{array}\right)\hat{\mathcal{T}}^{-1}\bar{\beta}_{\mu_{K}}\cdots\bar{\beta}_{\mu_{2}}\bar{\beta}_{\mu_{1}}\left|\phi\right\rangle }\label{eq:}\end{equation}
and therefore the right hand side of this expression can be evaluated
by means of the contractions of Eq. (\ref{eq:Multiqp-contr}) by substituting
the matrix $\mathbb{T}$ representing $\hat{\mathcal{T}}^{-1}$ by
the matrix $\mathbb{T}\left(\begin{array}{c}
\mu\\
\nu\end{array}\right)$ representing $\hat{\mathcal{T}}\left(\begin{array}{c}
\mu\\
\nu\end{array}\right)\hat{\mathcal{T}}^{-1}$. The existence of $\hat{\mathcal{T}}\left(\begin{array}{c}
\mu\\
\nu\end{array}\right)$, its explicit form as well as the expression of the matrix representing
this operator are considered thoroughly in Appendix C. As can be observed
in that appendix, the matrix representation of $\hat{\mathcal{T}}\left(\begin{array}{c}
\mu\\
\nu\end{array}\right)$ is nothing but a transposition matrix which exchanges given rows
or columns of the matrices applied to it. Therefore, the matrix $\mathbb{T}\left(\begin{array}{c}
\mu\\
\nu\end{array}\right)$ representative of $\hat{\mathcal{T}}\left(\begin{array}{c}
\mu\\
\nu\end{array}\right)\hat{\mathcal{T}}^{-1}$ can be obtained very easily out of the one representative of $\mathcal{T}^{-1}$
by exchanging the appropriate rows or columns.

\section{Conclusions}

By considering a certain limit of the statistical density operator
where it becomes the one of a pure state we have been able to obtain
the generalized Wick's theorem out of the corresponding statistical
version (Gaudin's theorem). The advantage of deriving the Generalized
Wick's Theorem (GWT) in this way is because the statistical version
is much easier to derive and handle as it is based on the cyclic property
of the trace over the Fock space. Using the limiting procedure we
have been able to obtain the most general contractions needed in the
GWT in a very easy way. By generalizing the limiting procedure we
have also been able to obtain a GWT for multiquasiparticle overlaps.
The corresponding contractions are given by simple and compact expressions
that can accommodate easily many quasiparticle excitations reducing
in this way the combinatorial complexity of the evaluation of multiquasiparticle
overlaps to just a degree of complexity associated to the number of
bodies of the operator whose multiquasiparticle overlap is required.
The expressions obtained for the overlap are new and they will be
very helpful in reducing the complexity of forthcoming beyond mean
field calculation. 

\begin{acknowledgments}
This work was supported in part by DGI, Ministerio de Ciencia y Tecnología,
Spain, under Project FIS2004-06697. S. P-M. acknowledges a scholarship
of the Programa de Formación del Profesorado Universitario (Ref. AP
2001-0182). 
\end{acknowledgments}
\appendix

\section{The exponential of one body operators as canonical transformation
operators}

In this appendix we will remind the reader about the expression of
$\exp(\hat{K})\alpha_{\rho}\exp(-\hat{K})$ where $\hat{K}=\frac{1}{2}\sum_{\mu\nu}\mathbb{K}_{\mu\nu}\alpha_{\mu}\alpha_{\nu}$
is an one body operator written in terms of the matrix $\mathbb{K}$
and the fermionic operators of the condensed notation. Using the canonical
anticommutation relations $\{\alpha_{\mu},\alpha_{\nu}\}=\sigma_{\mu\nu}$
is very easy to verify that $[\hat{K},\alpha_{\rho}]=-\sum_{\nu}(\sigma\mathbb{K}_{A})_{\rho\nu}\alpha_{\nu}$
where the skew-symmetric matrix $\mathbb{K}_{A}=\frac{1}{2}(\mathbb{K}-\mathbb{K}^{T})$
has been introduced. Using the same rules we obtain $[\hat{K}[\hat{K},\alpha_{\rho}]]=\sum_{\nu}(\sigma\mathbb{K}_{A})_{\rho\nu}^{2}\alpha_{\nu}$,
etc which allows us to finally write\begin{equation}
\exp(\hat{K})\alpha_{\rho}\exp(-\hat{K})=\alpha_{\rho}+[\hat{K},\alpha_{\rho}]+\frac{1}{2!}[\hat{K}[\hat{K},\alpha_{\rho}]]+\ldots=\sum_{\nu}(e^{-\sigma\mathbb{K}_{A}})_{\rho\nu}\alpha_{\nu}\label{eq:}\end{equation}
The matrix $\mathbb{M=}e^{-\sigma\mathbb{K}_{A}}$, owing to the skew-symmetric
character of $\mathbb{K}_{A}$, satisfies $\mathbb{M}^{T}=e^{\mathbb{K}_{A}\sigma}=\sigma e^{\sigma\mathbb{K}_{A}}\sigma$
or $\mathbb{M}\sigma\mathbb{M}^{T}=\sigma$ that is nothing but the
condition for the matrix $\mathbb{M}$ of being the matrix of a canonical
transformation.

Now we will show a result needed in the developments of the paper
that states that if the matrices $\mathbb{M}$ and $\mathbb{T}$ both
satisfy a relation of the type $\mathbb{M}\sigma\mathbb{M}^{T}=\sigma$
(i.e. they are matrices representing canonical transformations) then
the relation \begin{equation}
(1+\mathbb{M}\mathbb{T})^{-1}\sigma+\sigma(1+\mathbb{T}^{T}\mathbb{M}^{T})^{-1}=\sigma\label{eq:AppA_rel}\end{equation}
holds. We start by considering \begin{equation}
\sigma(1+\mathbb{T}^{T}\mathbb{M}^{T})^{-1}=\sigma(1-\mathbb{T}^{T}\mathbb{M}^{T}+\ldots)=\sigma-\mathbb{T}^{-1}\sigma\mathbb{M}^{T}+\ldots=\sigma-\mathbb{T}^{-1}\mathbb{M}^{-1}\sigma+\ldots=(1+\mathbb{T}^{-1}\mathbb{M}^{-1})^{-1}\sigma\label{eq:}\end{equation}
The right hand side can be written as $(1+\mathbb{M}\mathbb{T})^{-1}\mathbb{M}\mathbb{T}\sigma=\sigma-(1+\mathbb{M}\mathbb{T})^{-1}\sigma$
and from here the sought result of Eq. (\ref{eq:AppA_rel}) easily
follows.

Finally, by means of an example, we will argue that not all possible
canonical transformations satisfying $\mathbb{M}\sigma\mathbb{M}^{T}=\sigma$
are necessarily given in terms of a skew-symmetric matrix $\mathbb{K}_{A}$
by the expression $\mathbb{M=}e^{-\sigma\mathbb{K}_{A}}$. To this
end just consider the case of a bidimensional configuration space.
In this case, the most general skew-symmetric matrix is given by $\mathbb{K}_{A}=\left(\begin{array}{cc}
0 & a\\
-a & 0\end{array}\right)$ where $a$ is a complex number and therefore $e^{-\sigma\mathbb{K}_{A}}=\left(\begin{array}{cc}
e^{a} & 0\\
0 & e^{-a}\end{array}\right)$. On the other hand, the most general bidimensional matrix $\mathbb{M}$
satisfying the canonical transformation condition is given by both
$\mathbb{M}=\left(\begin{array}{cc}
m & 0\\
0 & 1/m\end{array}\right)$ and $\mathbb{M}=\left(\begin{array}{cc}
0 & m\\
1/m & 0\end{array}\right)$. Obviously, the first matrix is of the $e^{-\sigma\mathbb{K}_{A}}$
kind but not the second one.

\section{Calculation of the derivative of certain trace}

In this appendix we will compute the following derivative\begin{equation}
\frac{d}{d\lambda}\mathrm{Tr}[\ln(1+\mathbb{M}\mathbb{T}(\lambda))]\label{eq:}\end{equation}
where both $\mathbb{T}(\lambda)=\exp(-\lambda\sigma\mathbb{S})$ and
$\mathbb{M}$ are matrices of respective canonical transformations
(i.e. $\mathbb{S}$ is an skew-symmetric matrix). The logarithm in
the previous expression has to be interpreted as the corresponding
Taylor series expansion\begin{equation}
\ln(1+x)=x-\frac{x^{2}}{2}+\frac{x^{3}}{3}-\frac{x^{4}}{4}+\ldots\label{eq:}\end{equation}
Now we have to consider the derivative of $\mathbb{M}\mathbb{T}(\lambda)$
with respect to $\lambda$\begin{equation}
\frac{d}{d\lambda}\left(\mathbb{M}\mathbb{T}(\lambda)\right)_{\rho\sigma}=-\left(\mathbb{M}\mathbb{T}(\lambda)\sigma\mathbb{S}\right)_{\rho\sigma}\label{eq:}\end{equation}
its square\begin{eqnarray}
\frac{d}{d\lambda}\left(\mathbb{M}\mathbb{T}(\lambda)\right)_{\rho\sigma}^{2} & = & \frac{d}{d\lambda}\sum_{\tau}\left(\left(\mathbb{M}\mathbb{T}(\lambda)\right)_{\rho\tau}\left(\mathbb{M}\mathbb{T}(\lambda)\right)_{\tau\sigma}\right)\label{eq:}\\
 & = & -\sum_{\tau}\left(\left(\mathbb{M}\mathbb{T}(\lambda)\sigma\mathbb{S}\right)_{\rho\tau}\left(\mathbb{M}\mathbb{T}(\lambda)\right)_{\tau\sigma}+\left(\mathbb{M}\mathbb{T}(\lambda)\right)_{\rho\tau}\left(\mathbb{M}\mathbb{T}(\lambda)\sigma\mathbb{S}\right)_{\tau\sigma}\right)\nonumber \\
 & = & -\left(\mathbb{M}\mathbb{T}(\lambda)\sigma\mathbb{S}\mathbb{M}\mathbb{T}(\lambda)+\mathbb{M}\mathbb{T}(\lambda)\mathbb{M}\mathbb{T}(\lambda)\sigma\mathbb{S}\right)_{\rho\sigma}\nonumber \end{eqnarray}
and higher powers\begin{equation}
\frac{d}{d\lambda}\left(\mathbb{M}\mathbb{T}(\lambda)\right)_{\rho\sigma}^{3}=-\left(\mathbb{M}\mathbb{T}(\lambda)\sigma\mathbb{S}\left(\mathbb{M}\mathbb{T}(\lambda)\right)^{2}+\mathbb{M}\mathbb{T}(\lambda)\mathbb{M}\mathbb{T}(\lambda)\sigma\mathbb{S}{\mathbb{M}\mathbb{T}(\lambda)+\left(\mathbb{M}\mathbb{T}(\lambda)\right)}^{2}\mathbb{M}\mathbb{T}(\lambda)\sigma\mathbb{S}\right)_{\rho\sigma}\label{eq:}\end{equation}
The generalization to higher powers is fairly simple and we only have
to be careful with the non commutativity of the matrices involved.
The results obtained so far for the derivatives are not very useful
due to the increasing number of terms but fortunately we only need
to consider its trace. Using the cyclic invariance property of the
trace we can rearrange all different matrices in the derivatives to
end up with a general and compact expression\begin{equation}
\frac{d}{d\lambda}\mathrm{Tr}\left[\left(\mathbb{M}\mathbb{T}(\lambda)\right)^{n}\right]=-n\mathrm{Tr}\left[\left(\mathbb{M}\mathbb{T}(\lambda)\right)^{n}\sigma\mathbb{S}\right]\label{eq:}\end{equation}
Using now this result we can write\begin{eqnarray}
\frac{d}{d\lambda}\mathrm{Tr}[\ln(1+\mathbb{M}\mathbb{T}(\lambda))] & = & \frac{d}{d\lambda}\mathrm{Tr\left[\left(\mathbb{M}\mathbb{T}(\lambda)\right)\right]}-\frac{1}{2}\frac{d}{d\lambda}\mathrm{Tr\left[\left(\mathbb{M}\mathbb{T}(\lambda)\right)^{2}\right]+\frac{1}{3}\frac{d}{d\lambda}\mathrm{Tr\left[\left(\mathbb{M}\mathbb{T}(\lambda)\right)^{3}\right]+\ldots}}\label{eq:}\\
 & = & -\mathrm{Tr}\left[\left(\mathbb{M}\mathbb{T}(\lambda)-\left(\mathbb{M}\mathbb{T}(\lambda)\right)^{2}+\left(\mathbb{M}\mathbb{T}(\lambda)\right)^{3}+\ldots\right)\sigma\mathbb{S}\right]\nonumber \\
 & = & -\mathrm{Tr}\left[\left(1-\left(1+\mathbb{M}\mathbb{T}(\lambda)\right)^{-1}\right)\sigma\mathbb{S}\right]\nonumber \end{eqnarray}
that leads to the final result\begin{equation}
\frac{d}{d\lambda}\mathrm{Tr}[\ln(1+\mathbb{M}\mathbb{T}(\lambda))]=-\mathrm{Tr}\left[\sigma\mathbb{S}\right]+\mathrm{Tr}\left[\left(1+\mathbb{M}\mathbb{T}(\lambda)\right)^{-1}\sigma\mathbb{S}\right]\label{eq:}\end{equation}

\section{The operator of index transformation $\hat{\mathcal{T}}\left(\protect\begin{array}{c}
\mu\protect\\
\nu\protect\end{array}\right)$}

The operator $\hat{\mathcal{T}}\left(\begin{array}{c}
\mu\\
\nu\end{array}\right)$ is defined is such a way that it transforms the quasiparticle annihilation
and creation operators with index $\mu$ into the ones with index
$\nu$ and viceversa, i.e. \begin{equation}
\beta_{\mu}=\hat{\mathcal{T}}\left(\begin{array}{c}
\mu\\
\nu\end{array}\right)\beta_{\nu}{\hat{\mathcal{T}}\left(\begin{array}{c}
\mu\\
\nu\end{array}\right)}^{-1}\label{eq:tmunu}\end{equation}
or\begin{equation}
\beta_{\mu}\hat{\mathcal{T}}\left(\begin{array}{c}
\mu\\
\nu\end{array}\right)=\hat{\mathcal{T}}\left(\begin{array}{c}
\mu\\
\nu\end{array}\right)\beta_{\nu}\label{eq:}\end{equation}
for the quasiparticle annihilation operators $\beta_{\mu}$ and $\beta_{\nu}$.
It also has to leave the other quasiparticle operators unchanged,
i.e. \begin{equation}
\beta_{\sigma}\hat{\mathcal{T}}\left(\begin{array}{c}
\mu\\
\nu\end{array}\right)=\hat{\mathcal{T}}\left(\begin{array}{c}
\mu\\
\nu\end{array}\right)\beta_{\sigma}\label{eq:tmunusig}\end{equation}
for any index $\sigma$ different from $\mu$ and $\nu$. Finally
we will impose a normalization that makes it to fulfill $\hat{\mathcal{T}}\left(\begin{array}{c}
\mu\\
\nu\end{array}\right)|\phi\rangle=|\phi\rangle$ where $|\phi\rangle$ is the vacuum of the quasiparticle annihilation
operators $\beta_{\mu}$. This normalization is not the natural one
$\hat{\mathcal{T}}\left(\begin{array}{c}
\mu\\
\nu\end{array}\right)|\phi\rangle=-|\phi\rangle$ (i.e. exchanging two fermionic quasiparticle states should lead to
a minus sign) but is more convenient as it makes unnecessary to keep
trace of the minus sign in the related expressions. However, if the
reader feels more comfortable with the later normalization just multiplying
by a minus sign the definition of $\mathcal{\hat{\mathcal{T}}\left(\begin{array}{c}
\mu\\
\nu\end{array}\right)}$ below is enough. Using the notation of previous sections, one can
write the requirement of Eq. (\ref{eq:tmunu}) and Eq. (\ref{eq:tmunusig})
as well as the ones corresponding to the creation operators as \begin{equation}
\alpha_{\rho}\hat{\mathcal{T}}\left(\begin{array}{c}
\mu\\
\nu\end{array}\right)=\sum_{\sigma}{\mathbb{T}\left(\begin{array}{c}
\mu\\
\nu\end{array}\right)}_{\rho\sigma}\hat{\mathcal{T}}\left(\begin{array}{c}
\mu\\
\nu\end{array}\right)\alpha_{\sigma}\label{eq:}\end{equation}
where the bipartite matrix $\mathbb{T}\left(\begin{array}{c}
\mu\\
\nu\end{array}\right)$ is block diagonal and given by\begin{equation}
\mathbb{\mathbb{T}\left(\begin{array}{c}
\mu\\
\nu\end{array}\right)}=\left(\begin{array}{cc}
\widetilde{T}(\mu\leftrightarrow\nu) & 0\\
0 & \widetilde{T}(\mu\leftrightarrow\nu)\end{array}\right)\label{eq:}\end{equation}
The matrix $\widetilde{T}(\mu\leftrightarrow\nu)$ is a permutation
matrix which has the structure\begin{equation}
\widetilde{T}(\mu\leftrightarrow\nu)=\begin{array}{c}
\\\\\\\nu\\
\\\\\\\mu\\
\\\\\\\end{array}\left(\begin{array}{ccccccccccc}
1 &  &  & \vdots &  &  &  & \vdots\\
 & \ddots &  & \vdots &  &  &  & \vdots\\
 &  & 1 & \vdots &  &  &  & \vdots\\
\cdots & \cdots & \cdots & 0 & \cdots & \cdots & \cdots & 1 & \cdots & \cdots & \cdots\\
 &  &  & \vdots & 1 &  &  & \vdots\\
 &  &  & \vdots &  & \ddots &  & \vdots\\
 &  &  & \vdots &  &  & 1 & \vdots\\
\cdots & \cdots & \cdots & 1 & \cdots & \cdots & \cdots & 0 & \cdots & \cdots & \cdots\\
 &  &  & \vdots &  &  &  & \vdots & 1\\
 &  &  & \vdots &  &  &  & \vdots &  & \ddots\\
 &  &  & \vdots &  &  &  & \vdots &  &  & 1\end{array}\right)\label{eq:}\end{equation}
with all the matrix elements not explicitly given equal to zero. The
matrix elements of $\widetilde{T}(\mu\leftrightarrow\nu)$ can be
written in a very compact form as ${\widetilde{T}(\mu\leftrightarrow\nu)}_{ij}=\delta_{ij}+(\delta_{i\nu}-\delta_{i\mu})(\delta_{j\mu}-\delta_{j\nu})$.
This is a permutation matrix because, when applied to an arbitrary
matrix A to the right, it exchanges its $\mu$ and $\nu$ rows. On
the other hand, if the matrix is applied to the left then the columns
$\mu$ and $\nu$ of $A$ are exchanged. The permutation matrix $\widetilde{T}(\mu\leftrightarrow\nu)$
is real, symmetric and idempotent (${\tilde{T}(\mu\leftrightarrow\nu)}^{2}=\openone$).
It is possible to write the permutation operator $\hat{\mathcal{T}}\left(\begin{array}{c}
\mu\\
\nu\end{array}\right)$ as the exponential of an one-body operator \begin{equation}
\hat{\mathcal{T}}\left(\begin{array}{c}
\mu\\
\nu\end{array}\right)=\exp(\hat{K}\left(\begin{array}{c}
\mu\\
\nu\end{array}\right))\label{eq:}\end{equation}
with the one body operator $\hat{K}\left(\begin{array}{c}
\mu\\
\nu\end{array}\right)=\frac{1}{2}\sum_{\mu\nu}{\mathbb{K}\left(\begin{array}{c}
\mu\\
\nu\end{array}\right)}_{\rho\sigma}\alpha_{\rho}\alpha_{\sigma}$. Using the general result \begin{equation}
\hat{\mathcal{T}}\alpha_{\mu}\hat{\mathcal{T}}^{-1}=\sum_{\nu}\mathbb{T}_{\mu\nu}\alpha_{\nu}\label{eq:}\end{equation}
with $\mathbb{T}=\exp(-\sigma\mathbb{K})$ and taking into account
that \begin{equation}
\left(\begin{array}{cc}
0 & 1\\
1 & 0\end{array}\right)=\exp\left[i\frac{\pi}{2}\left(\begin{array}{cc}
1 & -1\\
-1 & 1\end{array}\right)\right]\label{eq:}\end{equation}
a little of algebra leads to the final result\begin{equation}
\hat{\mathcal{T}}\left(\begin{array}{c}
\mu\\
\nu\end{array}\right)=\exp\left\{ -\frac{i\pi}{2}\left(\beta_{\mu}^{+}\beta_{\mu}+\beta_{\nu}^{+}\beta_{\nu}-\beta_{\mu}^{+}\beta_{\nu}-\beta_{\nu}^{+}\beta_{\mu}\right)\right\} =\exp\left\{ -\frac{i\pi}{2}\left(\beta_{\mu}^{+}-\beta_{\nu}^{+}\right)\left(\beta_{\mu}-\beta_{\nu}\right)\right\} \label{eq:tmunu2}\end{equation}
Just for simplicity we will assume in this appendix that the creation
operators are the hermitian conjugates of the corresponding annihilation
ones and use the standard representation $\beta_{\mu}^{+}$ for them.
This assumption is not crucial and all the results presented in this
appendix are independent of it. Now let us assume that we want to
evaluate the matrix element $\langle\phi|\beta_{\nu_{K}}\ldots\beta_{\nu_{1}}\hat{\mathcal{T}}_{C}\hat{O}\beta_{\mu_{1}}^{+}\ldots\beta_{\mu_{K}}^{+}|\phi\rangle$
where $\hat{\mathcal{T}}_{C}$ is an arbitrary operator carrying out
a canonical transformation. In order to apply the main results of
the paper we have first to transform $\langle\phi|\beta_{\nu_{K}}\ldots\beta_{\nu_{1}}$
into $\langle\phi|\beta_{\mu_{K}}\ldots\beta_{\mu_{1}}\hat{\mathcal{T}}\left(\begin{array}{ccc}
\mu_{1} & \ldots & \mu_{k}\\
\nu_{1} & \ldots & \nu_{k}\end{array}\right)$ by means of a transformation operator that in a first sight could
be thought to be expressed as the product of elementary transformation
operators \begin{equation}
\hat{\mathcal{T}}\left(\begin{array}{ccc}
\mu_{1} & \ldots & \mu_{k}\\
\nu_{1} & \ldots & \nu_{k}\end{array}\right)=\hat{\mathcal{T}}\left(\begin{array}{c}
\mu_{1}\\
\nu_{1}\end{array}\right)\hat{\mathcal{T}}\left(\begin{array}{c}
\mu_{2}\\
\nu_{2}\end{array}\right)\ldots\hat{\mathcal{T}}\left(\begin{array}{c}
\mu_{k}\\
\nu_{k}\end{array}\right)\label{eq:}\end{equation}
However, a little care is needed as in the process of moving $\mathcal{T}\left(\begin{array}{c}
\mu_{1}\\
\nu_{1}\end{array}\right)$ to the left \begin{equation}
\langle\phi|\beta_{\mu_{k}}\ldots\beta_{\mu_{1}}\hat{\mathcal{T}}\left(\begin{array}{c}
\mu_{1}\\
\nu_{1}\end{array}\right)=\langle\phi|\beta_{\mu_{k}}\ldots\beta_{\mu{}_{2}}\hat{\mathcal{T}}\left(\begin{array}{c}
\mu_{1}\\
\nu_{1}\end{array}\right)\beta_{\nu_{1}}=\langle\phi|\beta_{\mu_{k}}\ldots\beta_{\mu{}_{3}}\hat{\mathcal{T}}\left(\begin{array}{c}
\mu_{1}\\
\nu_{1}\end{array}\right)\beta_{\mu{}_{2}}\beta_{\nu_{1}}=\ldots\label{eq:}\end{equation}
it might happen that one of the remaining quasiparticle operators
have an index $\mu_{j}$ that coincides with $\nu_{1}$ and therefore
the transformation operator $\hat{\mathcal{T}}\left(\begin{array}{c}
\mu_{1}\\
\nu_{1}\end{array}\right)$ turns it into $\mu_{1}$ and this will definitively alter the rest
of the operators needed to carry out the transformation. An economical
way to eliminate this possibility is to {}``discard'' in the transformation
operator those indices which are equal in both the $\mu_{j}$ and
$\nu_{j}$ sets making it impossible that one of the elementary operators
is changing more than one index. The mathematical expression for the
operator would then be\begin{equation}
\hat{\mathcal{T}}\left(\begin{array}{ccc}
\mu_{1} & \ldots & \mu_{K}\\
\nu_{1} & \ldots & \nu_{K}\end{array}\right)=\prod_{r=1,\ldots K;\:\mu_{i}\ne\nu_{j}}\hat{\mathcal{T}}\left(\begin{array}{c}
\mu_{r}\\
\nu_{r}\end{array}\right)\label{eq:TmunuFin}\end{equation}
The order of the elementary operators is irrelevant as they commute
among themselves and the only remaining detail is that the final product
of quasiparticle operators $\beta_{\tilde{\nu}_{K}}\ldots\beta_{\tilde{\nu}_{1}}$
contains the same quasiparticle indices as $\beta_{\nu_{K}}\ldots\beta_{\nu_{1}}$
but not necessarily in the same order. As the annihilation quasiparticle
operators anticommute among themselves the reordering of the product
to bring it to the desired form will introduce an additional sign
that will be denoted $f_{\nu}$ and is given by minus one to the number
of transpositions needed to reorder the indices. This sign does not
show up in the evaluation of the overlap matrix element \begin{equation}
\frac{\langle\phi|\beta_{\nu_{K}}\ldots\beta_{\nu_{1}}\hat{\mathcal{T}}_{C}\hat{O}\beta_{\mu_{1}}^{+}\ldots\beta_{\mu_{K}}^{+}|\phi\rangle}{\langle\phi|\beta_{\nu_{K}}\ldots\beta_{\nu_{1}}\hat{\mathcal{T}}_{C}\beta_{\mu_{1}}^{+}\ldots\beta_{\mu_{K}}^{+}|\phi\rangle}=\frac{\langle\phi|\beta_{\mu_{K}}\ldots\beta_{\mu_{1}}\hat{\mathcal{T}}\left(\begin{array}{ccc}
\mu_{1} & \ldots & \mu_{K}\\
\nu_{1} & \ldots & \nu_{K}\end{array}\right)\hat{\mathcal{T}}_{C}\hat{O}\beta_{\mu_{1}}^{+}\ldots\beta_{\mu_{K}}^{+}|\phi\rangle}{\langle\phi|\beta_{\mu_{K}}\ldots\beta_{\mu_{1}}\hat{\mathcal{T}}\left(\begin{array}{ccc}
\mu_{1} & \ldots & \mu_{K}\\
\nu_{1} & \ldots & \nu_{K}\end{array}\right)\hat{\mathcal{T}}_{C}\beta_{\mu_{1}}^{+}\ldots\beta_{\mu_{K}}^{+}|\phi\rangle}\label{eq:}\end{equation}
as it is the same both in the numerator and the denominator. It has
to be kept in mind, however, that the sign has to be considered in
the evaluation of the norm overlap (the denominator of the above expression).

Finally, let us consider the combined effect of $\hat{\mathcal{T}}\left(\begin{array}{c}
\mathbf{\mu}\\
\mathbf{\nu}\end{array}\right)$ (we now go back to a more compact notation for the indices $\mu$
and $\nu$) and an operator $\hat{\mathcal{T}}_{C}$ carrying out
an arbitrary canonical transformation with transformation matrix $\mathbb{T}$
($\alpha_{\rho}\hat{\mathcal{T}}_{C}=\sum_{\sigma}\mathbb{T}_{\rho\sigma}\hat{\mathcal{T}}_{C}\alpha_{\sigma}$)
\begin{eqnarray}
\alpha_{\rho}\hat{\mathcal{T}}\left(\begin{array}{c}
\mathbf{\mu}\\
\mathbf{\nu}\end{array}\right)\hat{\mathcal{T}}_{C} & = & \sum_{\sigma}{\mathbb{T}\left(\begin{array}{c}
\mathbf{\mu}\\
\mathbf{\nu}\end{array}\right)}_{\rho\sigma}\hat{\mathcal{T}}\left(\begin{array}{c}
\mathbf{\mu}\\
\mathbf{\nu}\end{array}\right)\alpha_{\sigma}\hat{\mathcal{T}}_{C}\nonumber \\
 & = & \sum_{\sigma\sigma'}{\mathbb{T}\left(\begin{array}{c}
\mathbf{\mu}\\
\mathbf{\nu}\end{array}\right)}_{\rho\sigma}\mathbb{T}_{\sigma\sigma'}\hat{\mathcal{T}}\left(\begin{array}{c}
\mathbf{\mu}\\
\mathbf{\nu}\end{array}\right)\hat{\mathcal{T}}_{C}\alpha_{\sigma'}\label{eq:}\\
 & = & \sum_{\sigma}\tilde{\mathbb{T}}_{\rho\sigma}\hat{\mathcal{T}}\left(\begin{array}{c}
\mathbf{\mu}\\
\mathbf{\nu}\end{array}\right)\hat{\mathcal{T}}_{C}\alpha_{\sigma}\nonumber \end{eqnarray}
with\begin{equation}
\tilde{\mathbb{T}}_{\rho\sigma}=\sum_{\sigma'}{\mathbb{T}\left(\begin{array}{c}
\mathbf{\mu}\\
\mathbf{\nu}\end{array}\right)}_{\rho\sigma'}\mathbb{T}_{\sigma'\sigma}\label{eq:}\end{equation}
Taking into account the decomposition of $\hat{\mathcal{T}}\left(\begin{array}{c}
\mathbf{\mu}\\
\mathbf{\nu}\end{array}\right)$ as the product of elementary transformations of Eq. (\ref{eq:TmunuFin})
we can finally write\begin{equation}
\tilde{\mathbb{T}}=\left(\prod_{r=1,\ldots K;\:\mu_{i}\ne\nu_{j}}\mathbb{T}\left(\begin{array}{c}
\mu_{r}\\
\nu_{r}\end{array}\right)\right)\mathbb{T}\label{eq:}\end{equation}
or in other words, the matrix $\tilde{\mathbb{T}}$ is obtained from
the matrix $\mathbb{T}$ by the exchange of the rows $\nu_{r}$ into
the $\mu_{r}$ ones. As an illustration of the procedure consider
the transformation of $\beta_{5}\beta_{9}\beta_{15}\beta_{2}\beta_{6}\beta_{4}\beta_{21}$
into $\beta_{19}\beta_{13}\beta_{6}\beta_{22}\beta_{34}\beta_{15}\beta_{1}$.
The repeated indices are 6 and 15 and they will be discarded. Then
the index 5 will turn into 19, the 9 into 13, the 2 into 22, the four
into 34 and finally the 21 into 1. In this way we will end up with
$\beta_{19}\beta_{13}\beta_{15}\beta_{22}\beta_{6}\beta_{34}\beta_{1}$
and we will need four transpositions to bring it into the desired
order (that is, $f_{\nu}=+1$). The matrix $\tilde{\mathbb{T}}$ will
be obtained out from the matrix $\mathbb{T}$ by exchanging its row
number 5 with its row number 19; its row number 9 with its row number
13 and so one. The whole procedure can be very easily implemented
on a computer procedure.

\end{document}